\documentclass[aps,10pt,longbibliography,superscriptaddress]{revtex4}
\usepackage{graphicx}
\usepackage{epstopdf}
\usepackage{amssymb}
\usepackage{mathrsfs}
\usepackage{amsmath}
\usepackage{ulem}
\usepackage{color}
\usepackage{latexsym}
\usepackage{amsfonts}
\usepackage{hyperref}
\usepackage{subfigure}
\usepackage{wasysym}
\usepackage{dcolumn}
\usepackage{bm}
\usepackage[percent]{overpic}
\usepackage{natbib}
\usepackage{booktabs}
\usepackage{gensymb}
\usepackage{setspace}
\begin{document}

\selectfont
\title{ {\it In silico} synthesis of microgel particles}

\author{Nicoletta Gnan}\email[]{nicoletta.gnan@roma1.infn.it}
\affiliation{CNR-ISC, Uos Sapienza, Piazzale A. Moro 2, 00185 Roma, Italy}
\affiliation{Dipartimento di Fisica, {\em Sapienza} Universit\`a di Roma, Piazzale A. Moro 2, 00185 Roma, Italy}

\author{Lorenzo Rovigatti}\email[]{lorenzo.rovigatti@uniroma1.it}
\affiliation{CNR-ISC, Uos Sapienza, Piazzale A. Moro 2, 00185 Roma, Italy}
\affiliation{Dipartimento di Fisica, {\em Sapienza} Universit\`a di Roma, Piazzale A. Moro 2, 00185 Roma, Italy}

\author{Maxime Bergman}
\affiliation{Physical Chemistry, Department of Chemistry, Lund University, Lund, Sweden}

\author{Emanuela Zaccarelli}\email[]{emanuela.zaccarelli@cnr.it}
\affiliation{CNR-ISC, Uos Sapienza, Piazzale A. Moro 2, 00185 Roma, Italy}
\affiliation{Dipartimento di Fisica, {\em Sapienza} Universit\`a di Roma, Piazzale A. Moro 2, 00185 Roma, Italy}

\definecolor{corr14okt}{rgb}{0,0,1} % blue
\definecolor{moved}{rgb}{0,0,0}

\begin{abstract} 
Microgels are colloidal-scale particles individually made of crosslinked polymer networks that can swell and deswell in response to external stimuli, such as changes to temperature or pH. Despite a large amount of experimental activities on microgels, a proper theoretical description based on individual particle properties is still missing due to the complexity of the particles. To go one step further, here we propose a novel methodology to assemble realistic microgel particles {\it in silico}. We exploit the self-assembly of a binary mixture composed of tetravalent (crosslinkers) and bivalent (monomer beads) patchy particles under spherical confinement in order to produce fully-bonded networks. The resulting structure is then used to generate the initial microgel configuration, which is subsequently simulated with a bead-spring model complemented by a temperature-induced hydrophobic attraction. To validate our assembly protocol we focus on a small microgel test-case and show that we can reproduce the experimental swelling curve by appropriately tuning the confining sphere radius, something that would not be possible with less sophisticated assembly methodologies, \textit{e.g.} in the case of networks generated from an underlying crystal structure. We further investigate the structure (in reciprocal and real space) and the swelling curves of microgels as a function of temperature, finding that our results are well described by the widely-used fuzzy sphere model. %{\bf (and Flory-Huggins theory??).} 
This is a first step toward a realistic modelling of microgel particles, which will pave the way for a careful assessment of their elastic properties and effective interactions.
\end{abstract} 

\maketitle
\section{Introduction}
Soft colloids, combining properties of hard-sphere colloids and polymers, offer the interesting possibility to tailor their macroscopic behaviour and flow at the molecular level\cite{vlassopoulos2014tunable}.  Hard sphere colloids have served for decades as a reference model to shed light on many physics problems, such as the structure of atomic liquids\cite{hansen2013theory}, crystal nucleation\cite{auer2001prediction} and the glass transition\cite{pusey1986phase,brambilla2009probing,zaccarelli2015polydispersity}. Recently, however, soft colloids have become even more popular in the scientific community\cite{yunker2014physics}. Typically, soft particles have an internal polymeric architecture, which allows them to reach the paste regime\cite{seth2006elastic}, where particles are in very dense, squeezed (or jammed) states. Under such conditions, the polymeric chains experience a substantial interpenetration\cite{mohanty2017interpenetration}, and the internal structure of particles themselves changes. Since the single-particle elasticity alters the properties of the resulting macroscopic material, establishing the crucial link between microscopic properties and macroscopic response requires the knowledge of the effective interactions among particles. Yet, while synthesis and experiments have produced an immense database of soft polymeric colloids\cite{vlassopoulos2014tunable}, theoretical efforts still lag behind. 
The most dramatic example of such a dichotomy is offered by microgels, which are soft particles with tunable swelling properties and an extremely broad range of materials applications. Microgels are nano- or microsized particles made by cross-linked polymer networks, which can respond to external stimuli, such as changes to temperature $T$ or pH, by swelling and deswelling (on a short time-scale as compared to macrogels). The important experimental advantage of using microgels as a model system is that, thanks to the aforementioned responsivity, the size of the particles can be finely controlled. Such a sensitivity is usually exploited to carefully tune the sample volume fraction without changing the particle number density. This property allows to explore classical physics problems from new points of view: for example, by using confocal microscopy, it is possible to follow the melting of colloidal crystals by tuning the temperature, allowing to investigate fluidization events at grain boundaries\cite{alsayed2005premelting} or to study the nature of the fluid-hexatic-crystal transition in 2D\cite{han2008melting}. Similarly, a slow annealing induces a reversible assembly of crystals even in the presence of defects\cite{iyer2009self}, while size changes can be used to smartly vary the lattice spacing and to obtain color-tunable crystals\cite{debord2002color}. 

Among microgels, largely studied are those based on PNIPAM crosslinked networks, which are thermoresponsive and display a Volume Phase Transition (VPT) at $T\sim 32\degree$~C, from a swollen state at low $T$ to a compact state at high $T$. The first synthesis of PNIPAM microgels in 1986\cite{pelton1986preparation} was followed by hundreds of publications, including several reviews on synthesis\cite{saunders1999microgel,pelton2000temperature}, applications\cite{oh2008development,fernandez2009gels} and physical aspects \cite{lyon2012polymer,yunker2014physics} as well as several books\cite{pich2010microgels,fernandez2011microgel,sadowski2014intelligent}.
Despite microgels being one of the most experimentally studied soft matter systems, most theoretical efforts have adopted so far an approach that neglects the polymeric nature of the particles, even though at high density this aspect becomes overwhelmingly important. 
As underlined in a recent review by Lyon and Fernandez-Nieves\cite{lyon2012polymer}, the polymer/colloid duality of microgels, which is at the core of their technological and fundamental relevance, has largely been overlooked. Thus, the description of microgel phase behavior and dynamics has mostly been based on coarse-grained effective interactions\cite{likos2001effective} as simple as the Hertzian model for elastic spheres\cite{zhang2009thermal,pamies2009phase,mohanty2014effective}.
Only in the case of ionic microgels an effective Hamiltonian has been derived from first-principles, thanks to the fact that electrostatics is the dominant contribution\cite{denton2003counterion,likos2011structure}. 

The purpose of this work is to build up a flexible numerical protocol able to design individual microgel particles {\it  in silico} with properties comparable to the experimental ones. It is worth stressing that we do not seek to realistically reproduce the kinetics of the chemical synthesis, but rather to deliver a final product with characteristics as much as possible to the synthesised particles. In this work we will thus build the microgel particles and compare their swelling behaviour with experimental results. To do so, we specifically focus on the case of very small microgels, or nanogels, whose diameter in the swollen regime is approximately $\sim 50\,$nm. This value appears to be close to the smallest possible size for which stable microgel particles can be synthesized using surfactant-based methods\cite{pelton2004unresolved}. We choose to work with such small microgels because in this case we are able to reproduce the network in a monomer-resolved way by using the classic bead-spring model for polymers. In this respect, each bead has a size comparable to the Kuhn length of the PNIPAM chains\cite{rubinstein2010polymer}, thus avoiding complications due to coarse-graining.

The strategy we devise makes use of a binary mixture of patchy particles under spherical confinement to self-assemble fully-bonded disordered networks. These networks are then topologically constrained and simulated using classical polymer models. By contrast, previous numerical efforts in microgel modelling have focused on unrealistic networks formed by chains of the same length, often connecting crosslinkers placed on crystalline lattice sites\cite{claudio2009comparison,jha2011study,kobayashi2014structure,ghavami2016internal,Ahuali2017,kobayashi2017polymer}. A notable exception is provided in Refs.~\cite{elvingson_jpcm,elvingson_collapse}, where closed polymer networks are constructed by directly joining randomly-distributed crosslinkers with polymer chains.

In addition to their disordered nature, the networks we generate possess the additional advantage of having their swelling properties dependent on the extent of the confinement employed during the initial assembly. We find an excellent agreement between the measured swelling curve for small microgels and our numerical results upon rather strong confinement. At the same time, we show that the experimental behaviour cannot be reproduced using a crystal-based network microgel of the same size. We further investigate the internal structure of the particles and successfully compare the form factors with the widely used fuzzy sphere model~\cite{fernandez2011microgel}, thus achieving a complete characterization of the particles. Our study is thus a first step toward a realistic microgel modelling which fully incorporates and takes into account the polymeric nature of the particles.%\cite{lyon2012polymer}. 
%In future work, aiming provide answers to several of the open questions on microgels This leaves open a number of questions, such as: How do microgels behave at high packing fractions? When do they strongly interpenetrate? What is the structure of microgel surface under very dense conditions and how does this reflect on microgel-microgel effective interactions? Can these microscopic features be related to the consequent change of macroscopic behaviour and mechanical properties of bulk suspensions?

\section{Materials and Methods}

\subsection{Microgel synthesis and experimental characterization}
We specifically focus on the modellization of PNIPAM microgels synthesised \textit{via} precipitation polymerisation through the combination of
$0.1929\,$g of sodium dodecyl sulphate (SDS, Duchefa Biochemie), $1.471\,g$ of NIPAM and $0.0647\,$g of BIS in $96.29\,$g water. NIPAM was re-crystallised in hexane and all other chemicals were used as received. The mixture was heated to 70$\degree$C, bubbled with argon and $0.0539\,$g of KPS in $2.0145\,$g of water was added to start the reaction. The reaction was then left for 6 hours under argon atmosphere. The particle suspensions were cleaned by three centrifugation and re-dispersion series. The crosslinker concentration is thus 3.2 mol\% with respect to NIPAM. 

The characterization of the particles was carried out through dynamic light scattering (DLS) in pseudo-2D cross-correlation mode with laser wavelength $\lambda = 660\,$nm (LS instruments, Switzerland). DLS measurements were performed over a range of temperature in steps of $2\,$K to yield an accurate swelling curve. The hydrodynamic radius was extracted using a first order cumulant analysis averaged over an angular range of 60-100$\degree$ every 10$\degree$. The microgels have a hydrodynamic radius $R_H \sim 24$~nm under the most swollen investigated conditions ($T=282K$).

\subsection{Design, Assembly of the network and swelling simulations}

\begin{figure}[h!]
\includegraphics[width=0.7\textwidth]{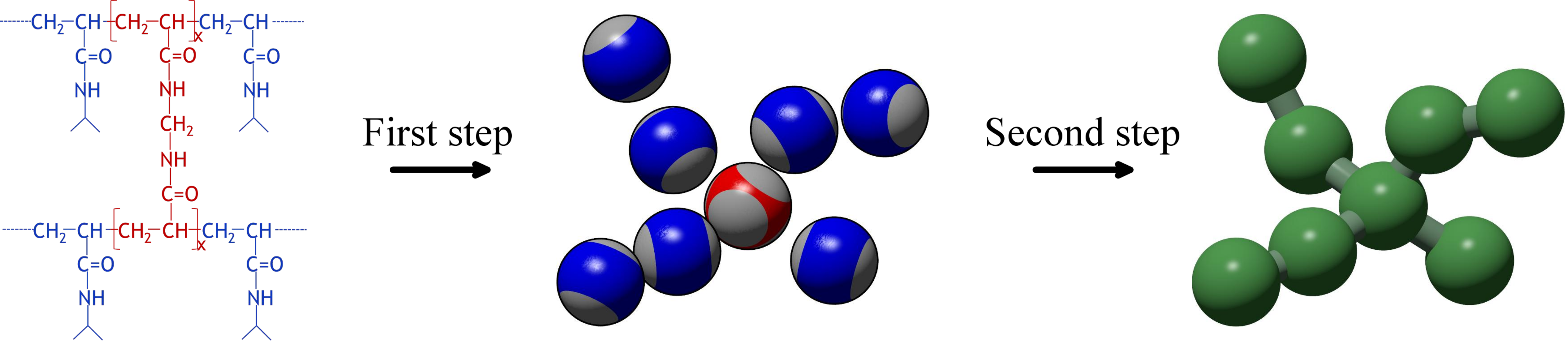}
\caption{\label{fig:sketch} A sketch of the procedure employed in this work to synthesise \textit{in silico} microgels. The monomer and crosslink molecules (leftmost panel) are initially modelled as patchy particles with two and four patches, respectively. This is shown in the central panel, where bivalent patchy particles (mimicking NIPAM monomers) are coloured in blue, tetravalent ones, acting as BIS crosslinkers, in red, and patches are represented in grey. After the network has assembled, the resulting configuration is used as a starting point for simulations performed with a bead-spring model (rightmost panel).}
\end{figure}

To build the initial microgel configuration we consider a mixture of bivalent and tetravalent patchy particles under spherical confinement. Each patchy particle is modelled as a sphere whose surface is decorated by either four patches of species A arranged on a tetrahedron or two patches of species B placed on the poles. A patch $\mu$ on particle $i$ is identified by the unit vector $\mathbf{p}^\mu_i$. The two-body interaction potential between particles $i$ and $j$ reads
\begin{equation}
\label{eq:two_body}
V(\mathbf{r}_{ij}, \lbrace\mathbf{p}_i\rbrace, \lbrace\mathbf{p}_j\rbrace) = V_{\rm WCA}(r_{ij}) + \sum_{\mu \in \lbrace p_i \rbrace} \sum_{\nu \in \lbrace p_j \rbrace} V_{\rm patchy}(r_{\mu\nu})
\end{equation}
\noindent
where $\mathbf{r}_{ij}$ is the vector connecting $i$ and $j$, $r_{ij}$ is its length, $\lbrace p_k \rbrace$ is the set of patches of particle $k$ and $r_{\mu\nu}$ is the distance between patch $\mu$ on particle $i$ and patch $\nu$ on particle $j$. The two contributions $V_{\rm WCA}$ and $V_{\rm patchy}$ encode a short-range repulsion and a short-range attraction, respectively. The repulsion is modelled with a Weeks-Chandler-Andersen potential~\cite{wca},
\begin{eqnarray}
V_{\rm WCA}(r) = 
\begin{cases}
4\epsilon\left[\left(\frac{\sigma}{r}\right)^{12} - \left(\frac{\sigma}{r}\right)^6 \right] + \epsilon & {\rm if} \quad r \leq 2^\frac{1}{6}\sigma \\[0.5em]
0 & {\rm if} \quad r > 2^\frac{1}{6}\sigma
\end{cases}
\label{eq:WCA}
\end{eqnarray}
\noindent
where $\sigma$ is the particle diameter, which is taken as the unit of length, and $\epsilon$ controls the energy scale. The patch-patch interaction reads
\begin{eqnarray}
V_{\rm patchy}(r_{\mu\nu}) = 
\begin{cases}
2\epsilon_{\mu\nu}\left(\frac{\sigma_{p}^4}{2r_{\mu\nu}^4} - 1\right)e^{\frac{\sigma_{p}}{(r_{\mu\nu} - r_c)} + 2} & {\rm if} \quad r_{\mu\nu} \leq r_c\\[0.5em]
0 & {\rm if} \quad r_{\mu\nu} > r_c
\end{cases}
\label{eq:Vpatchy}
\end{eqnarray}
where $\sigma_p$ sets the position of the attractive well (of depth $\epsilon_{\mu\nu}$) and $r_c$ is chosen by imposing $V_{\rm patchy}(r_c) = 0$. Here we set $\sigma_p = 0.4$, $r_c = 1.5\sigma_p$. In addition, we set $\epsilon_{AB} = \epsilon_{BB} = \epsilon$ and $\epsilon_{AA} = 0$, so that only bonding between BB and AB patches is possible. 
The two-body interaction given in Eq.~\eqref{eq:two_body} is complemented by a three-body potential acting on triplets of close patches\cite{Sciortino2017}.
This term provides an efficient bond-swapping mechanism that makes it possible to easily equilibrate the system at extremely low temperatures, while, at the same time, the single-bond-per-patch condition. Additional details can be found in Ref.~\cite{Sciortino2017}.

We focus on systems of $N_A$ tetravalent particles, acting as crosslinkers, and $N_B$ bivalent particles, mimicking the NIPAM monomers. We set $N_A + N_B = 42000$  and $N_A=0.032 (N_A+N_B)=1344$ to impose the same crosslinker concentration as in experiments. Each monomer corresponds to a bead of size compatible with the Kuhn length\cite{rubinstein2010polymer} of PNIPAM polymers\cite{kutnyanszky2012there,magerl2015influence}, i.e. $\sigma \sim 1$nm. Thus the total number of monomers employed for the assembly has been estimated by imposing that the volume of the microgel particles is the same in experiments and simulations. 
We perform Molecular Dynamics (MD) simulations at fixed temperature $T^* = k_BT/\epsilon = 0.05$, where $k_B$ is the Boltzmann constant. Thanks to such a low temperature, the system tends to maximize the number of bonds. In addition, owing to the bond-swapping mechanism, the system is able to continuously restructure itself, until the large majority of possible bonds are formed. 

To ensure that the resulting network retains the spherical shape of a microgel colloid, we confine particles in a sphere of radius $Z$, with no periodic boundary conditions applied. The simulations are left running until the great majority of the particles ($95$ -- $99\%$, depending on the system parameters) has self-assembled into a single network-like cluster. The assembled microgels contain a negligible fraction of dangling ends, since nearly all the bivalent particles belonging to the largest cluster are involved in two bonds. Indeed, the ratio between the number of formed bonds and the total number of possible bonds is always larger than 0.998.  This assembly stage takes from a few hours to a few days on K80 NVIDIA GPUs, depending on crosslinker concentration and confinement. In particular, the aggregation rate is slower the fewer the crosslinkers or the larger the confinement. At the end of the procedure, we remove all those particles that are not in the largest cluster and use the resulting structure (typically comprising $N\sim 41000$ particles) as the initial configuration for the bead-spring model simulations.

Once the network is formed, we preserve its topology by replacing the directional interactions among the particles with a standard set of interactions designed to reproduce the behaviour of polymers~\cite{kremer1990dynamics}. To mimic the interactions between polymers, we adopt the classic bead-spring model in which bonded monomers $i$ and $j$ interact through the sum of a WCA potential (the same of Eq.~\eqref{eq:WCA}) and a FENE potential $V_{\rm FENE}$~\cite{kremer1990dynamics, bernabei2011chain, soddemann2001generic}:
\begin{eqnarray}
\label{eq:V_FENE}
V_{\rm FENE}(r)=-\varepsilon \, k_{F}R_0^2 \ln(1-(\frac{r}{R_0\sigma})^2) & {\rm if} \quad  r< R_0\sigma \\[0.5em]
\end{eqnarray}
\noindent
where $k_F=15$ is the spring constant and $R_{0}=1.5$  is the maximum extension of the bond. In this way the reversible bonds formed by patchy particles are replaced by permanent bonds between connected monomers which preserve the initial network topology at all times.
wh-bonded monomers interact only through the WCA potential in Eq.~\eqref{eq:WCA}. 

In order to grasp the swelling behavior of PNIPAM microgels with temperature, we need to take into account the tunable quality of the solvent. We do so in an implicit way, by adding a term $V_{\alpha}$ to the interactions between all monomer pairs. The strength of this term is set by the parameter $\alpha$, which controls the solvophobicity of the monomers and plays the role of a temperature~\cite{soddemann2001generic,verso2015simulation}. This term reads,
\begin{eqnarray}
\label{eq:V_alpha}
V_{\alpha}(r) =
\begin{cases}
-\varepsilon\alpha & {\rm if} \quad r \leq 2^{1/6} \sigma\\[0.5em]
\frac{1}{2}\alpha\varepsilon [\cos(\gamma (r/\sigma)^2+\beta) -1] & {\rm if} \quad 2^{1/6}\sigma < r\leq R_{0} \sigma\\[0.5em]
0 & {\rm otherwise}
\end{cases}
\end{eqnarray}
\noindent where we set $\gamma=\pi (2.25-2^{1/3})^{-1}$ and $\beta=2\pi-2.25\gamma$~\cite{soddemann2001generic}. In the $\alpha=0$ limit, the system is in good solvent conditions and no additional attractive interactions are at work.
We perform MD simulations of a single microgel particle at constant temperature $T^*=1.0$, using Nos\`{e}-Hoover thermostat and a leap-frog integration scheme with a time step $\delta t=0.001$, with the same units as described above. We increase $\alpha$ until a complete collapse of the particles is observed.

We note that, in the standard chemical synthesis \cite{pelton2011microgels}, the polymerization kinetics takes place at temperatures $\sim 70\degree$C to facilitate aggregation. It is well-known that, under these bad solvent conditions, the crosslinkers react faster than the NIPAM monomers, soon forming a rather homogeneous branched core, which later incorporates additional chains and smaller aggregates. Other components, such as surfactants, are often added to obtain particles with small sizes\cite{pelton2011microgels}. Other parameters, such as the initiator concentration\cite{virtanen2016persulfate}, may influence the synthesis and the final density profile in ways that are not fully understood yet\cite{elaissari2011polymerization,hoare2006kinetic}. %Indeed, no theoretical models that account for the whole set of chemical processes leading to the final products are available\cite{elaissari2011polymerization,hoare2006kinetic}. 
%Nonetheless, the whole process takes several minutes (and thus rearrangements are possible? sono solo legami chimici o no?)
Therefore, we stress that in our approach we do not aim to reproduce the experimental protocol for the synthesis in its different aspects, but rather to adjust the several involved parameters, such as the confinement but also the patchy potential between different species, in order to achieve network realizations that are as similar as possible to the experimental particles in their basic properties, such as swelling behavior and elastic properties. The latter will be addressed in future work.

Finally we compare the structural properties of our microgels with those obtained from a microgel generated out of an ordered network based on a diamond lattice, as recently done in Refs.~\cite{jha2011study,kobayashi2014structure,ghavami2016internal,kobayashi2017polymer}. In this approach the sites of the lattice represent the crosslinkers and each two sites are connected by a polymer chain. The spherical shape of the microgel is then obtained by cutting the network along a spherical surface. By construction, all chains have the same number of monomers. Consequently, the size of the microgel cannot be tuned by maintaining the same total number of particles. In the following we will employ a diamond-lattice-based microgel with $N~42000$ monomers and $c=3.2\%$.

\section{Results}

\subsection{Swollen regime}

We generate microgel configurations for different values of the confinement radius in the range  $30\sigma < Z < 70\sigma$. For each value, we replicate the assembly protocol a few times $2$ -- $4$ independent realizations are already enough to suppress any significant numerical noise) in order to average properties over independent conformations. After a fully bonded network is obtained, the patchy interactions are replaced by the bead-spring ones (with $\alpha=0$) until equilibration is obtained.
\begin{figure}[h!]
\includegraphics[width=0.7\textwidth]{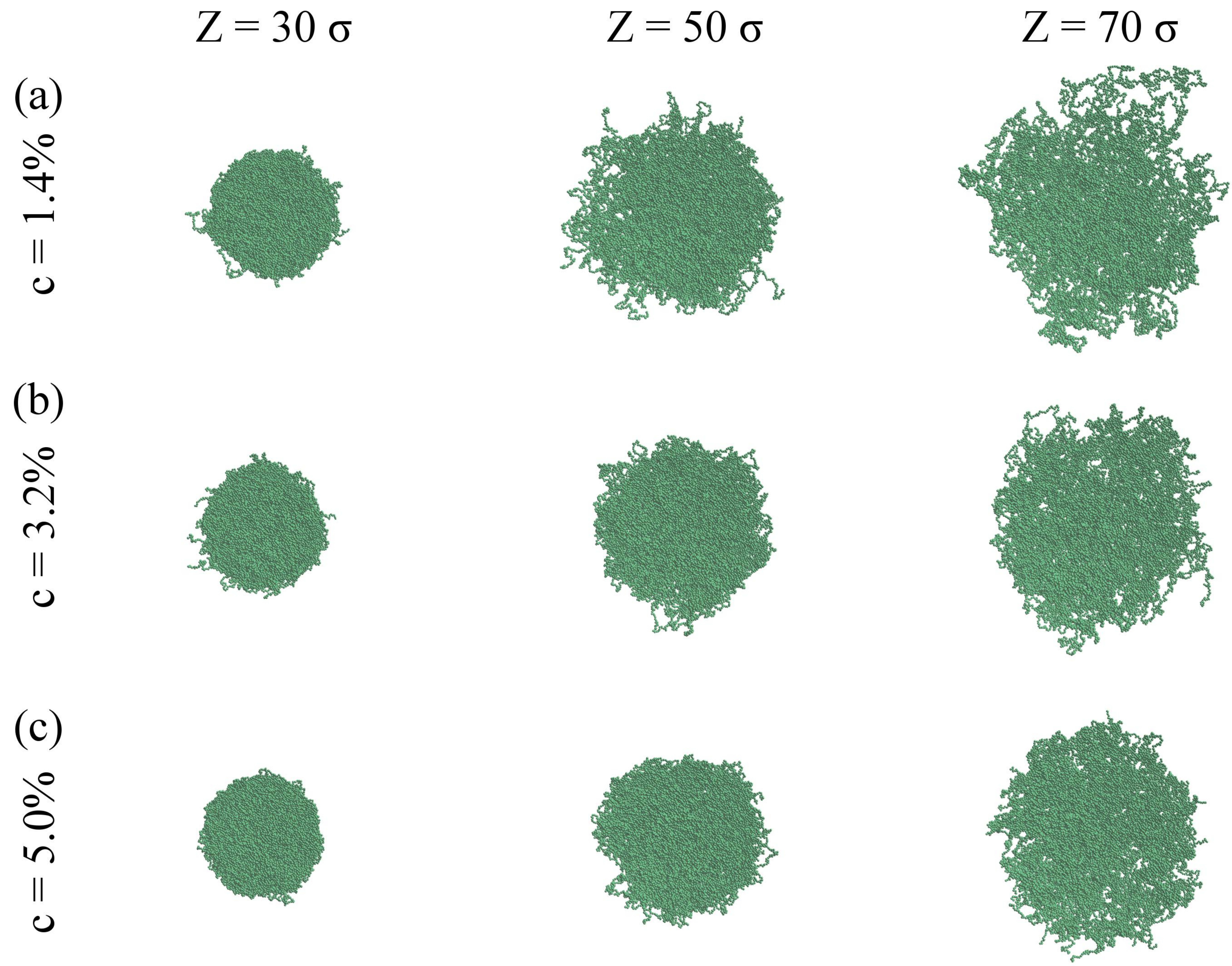}
\caption{\label{fig:snapshots}Equilibrated simulation snapshots of microgels generated at (a) $c=1.4\%$, (b) $c=3.2\%$ and (c) $c=5.0\%$ and three different confinements: from left to right, $Z=30\sigma$, $50\sigma$ and $70\sigma$.}
\end{figure}

Figure~\ref{fig:snapshots} shows typical equilibrated snapshots of the microgels for different choices of crosslinker concentration $c$ and $Z$. Immediately we notice that the qualitative effect of reducing the number of crosslinkers is the same as that of increasing the confinement radius. In both cases the microgel tends to be larger and less compact, even though the properties of the gel networks should be quantitatively different. However, the reduction of the confining volume has a more dramatic effect on the size of the microgels than the increase of the number of crosslinkers for the range of parameters considered here. In addition, as the confining radius gets larger and larger, the conformation of the microgels becomes more irregular. Indeed, even though it retains on average a spherical shape, there are more and more portions of long chains that stick out of the corona before looping back towards the microgel core. In the framework of the simplified core-corona model often invoked to describe microgel behavior\cite{rey2016isostructural}, it is expected that the corona properties, particularly the elasticity and the effective interactions, will be largely affected by these outer chains. Indeed, recent work by Boon and Schurtenberger\cite{boon2017swelling} has shown that these need to be taken into account to fully describe the experimental density profiles in the corona. Our approach is thus able to incorporate this effect\cite{footnoteends}.

\begin{figure}[h!]
\includegraphics[width=0.48\textwidth,clip]{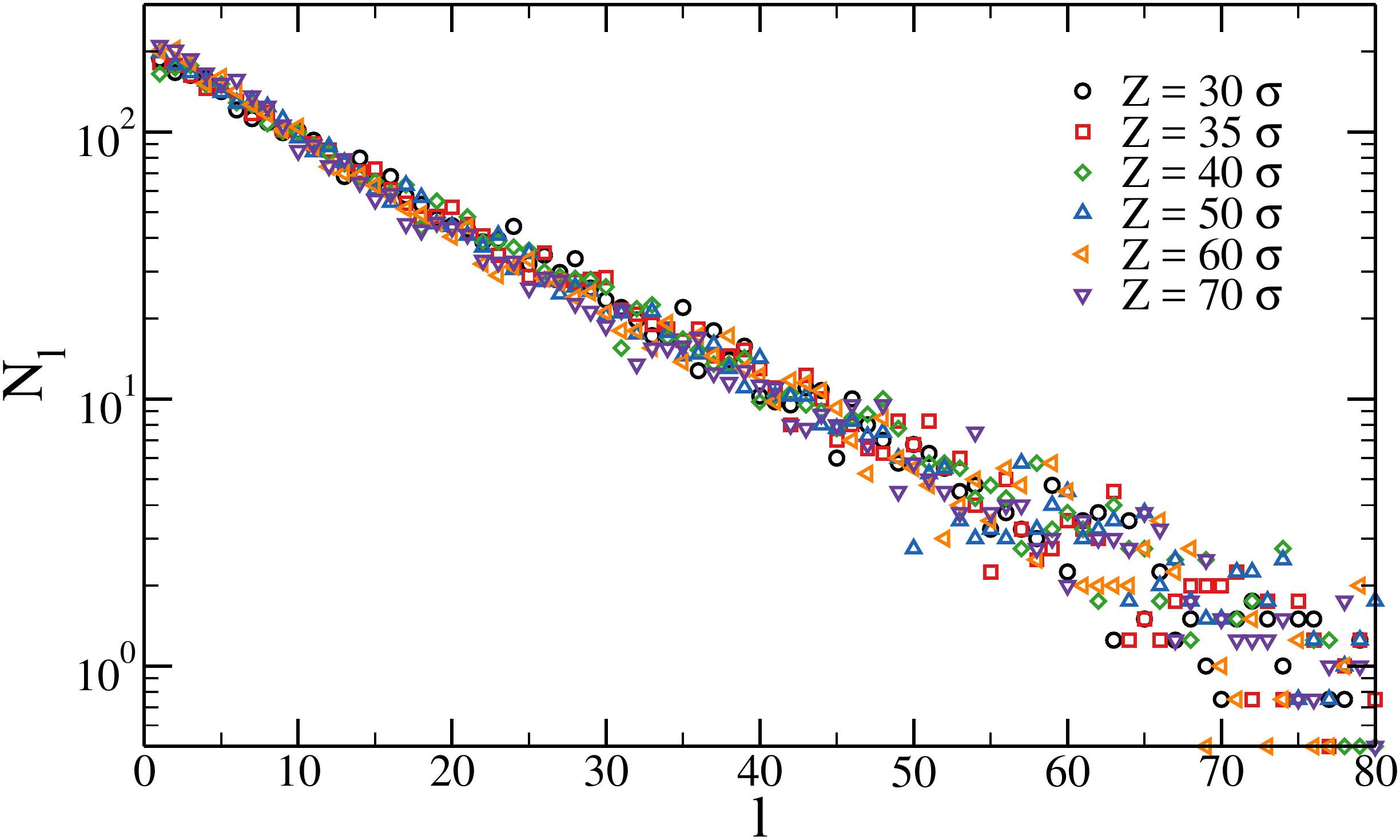}
\caption{\label{fig:chains_csd}Chain size distribution for $c=3.2\%$ microgels at different confinement radius $Z$.}
\end{figure}

Regarding the distribution of chain sizes inside the network, Figure~\ref{fig:chains_csd} shows $N_l$, the number of chains of size $l$,  which is defined as segments formed by $l$ bonded bivalent particles, as a function of confinement $Z$. Quite strikingly, the distribution is found to be always exponential and independent of the confinement.. This is a consequence of the equilibrium nature of the self-assembly process of the constituent patchy particles. Indeed, for bivalent particles it was found that $N_l$ is always exponential, provided that the system undergoes an equilibrium polymerisation~\cite{bianchi_FS}. Here, we show that this result holds even in the presence of crosslinkers and is not affected by the presence of the confinement.

\begin{figure}[h!]
\includegraphics[width=0.48\textwidth,clip]{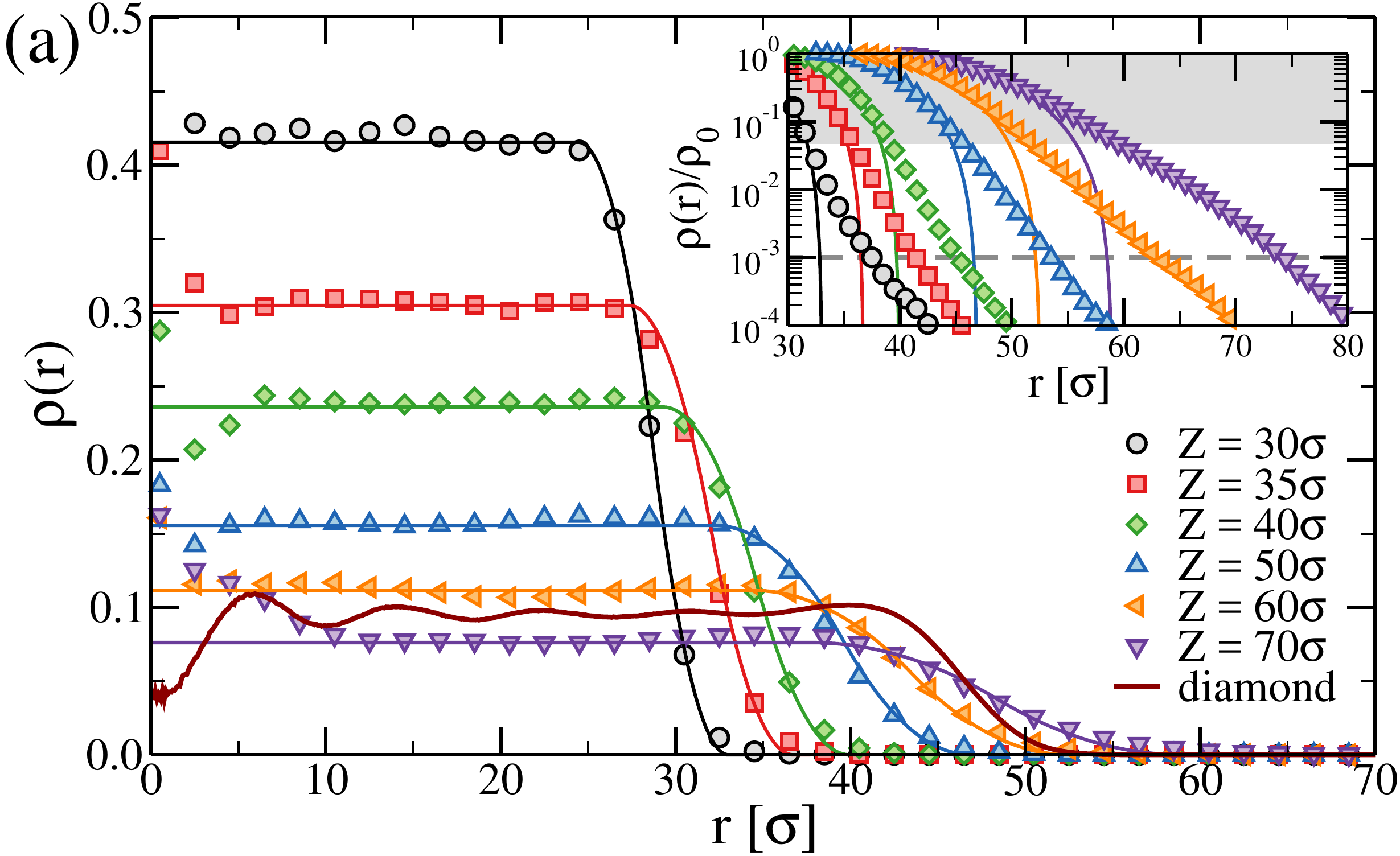}
\includegraphics[width=0.47\textwidth,clip]{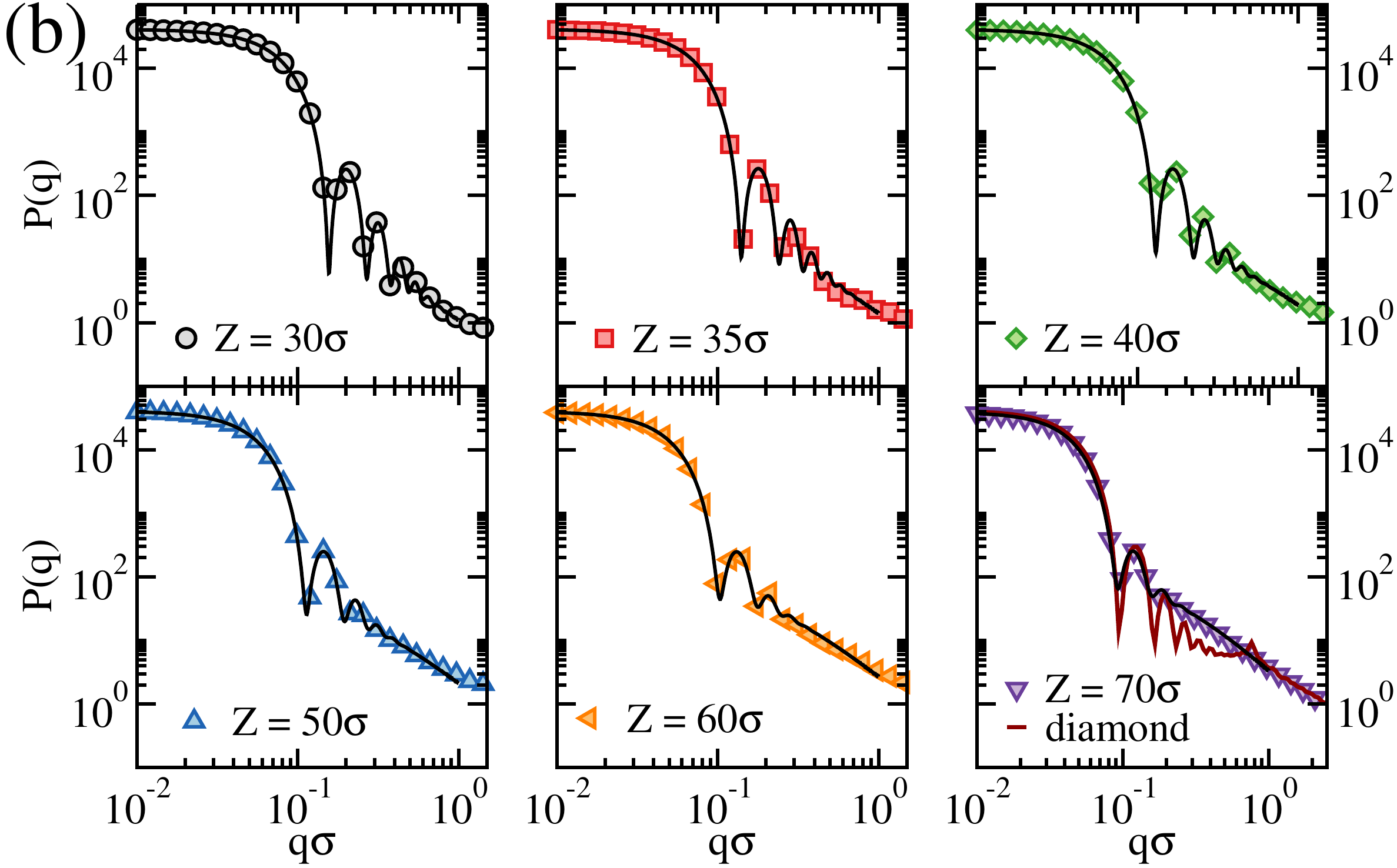}\\[1em]
\caption{\label{fig:profiles_3.2}(a) Monomer density profiles for $c=3.2\%$ microgels at different confinement radius $Z$ as a function of the distance $r$ from the microgel centre of mass. Here symbols are simulations results, lines are obtained from the fits of the form factors using the fuzzy sphere model and applying Eq.~\eqref{eq:profiles}. The inset shows the same data rescaled by the polymer density of the inner core, which changes with $Z$, on a log-linear scale. The shaded area shows the approximate extent of the region where the curves obtained through Eq.~\eqref{eq:profiles} well-reproduce the numerical data. The value of $r$ at which each dataset intersects the dashed grey line defined by $\rho(r)/\rho_0 \equiv 10^{-3}$ provides an estimate of the hydrodynamic radius, $R_H$. (b) Form factors $P(q)$ for microgels generated with different confinements. Symbols are simulation data, lines are fits to Eq.~\eqref{eq:fuzzy}. The (dark red) thick lines in (a) and in the bottom-rightmost panel of (b) refer to a microgel generated by a diamond lattice configuration.}
\end{figure}

From now on, we focus on a fixed crosslinker concentration $c=3.2\%$, which is the same used in the experiments reported in this work. First we calculate the monomer density profile $\rho(r)$ as a function of the distance $r$ from the center of the microgel. This is shown in Fig.~\ref{fig:profiles_3.2}(a) for several values of the confining radius. We see that, for all values of $Z$, we have a well-defined core, characterized by a homogeneous, flat profile, followed by a corona where the density decays to zero with a slope that depends on $Z$. Therefore, the choice of the confinement allows us to tune the extent of the core and the corona profile. It is thus a useful parameter that can be adjusted to compare with experiments. 

We also calculate the density profile of a diamond-lattice-based microgel, for which  it is possible to obtain only one density profile once the number of crosslinkers is fixed. This is shown in Fig.~\ref{fig:profiles_3.2} (a) for a diamond-lattice microgel with $N \approx 42000$ and $c=3.2\%$.
It is interesting to note that the density profile of the diamond-lattice microgel is characterized by oscillations of period $\approx 8 \sigma$ inside the core, a signature of the ordered polymer network.

Except for very recent super-resolution microscopy measurements\cite{conley2016superresolution}, typically
the density profiles in experiments are indirectly obtained from the form factors $P(q)$, which can be measured by neutron or x-ray scattering as a function of the wavenumber $q$. In general, microgel form factors are well-described by the fuzzy-sphere model~\cite{stieger2004small}, which assumes that the microgel particle has a homogeneous density profile in the core, being then surrounded by a corona of decreasing density. A Lorentzian function is also incorporated in the model to account for the network fluctuations due to the presence of static and dynamic inhomogeneities in the polymer network. Thus, the fit expression for the form factors is,
\begin{equation}\label{eq:fuzzy}
P(q)=\left [\frac{3 [ \sin(qR') -qR' \cos(qR')] }{(qR')^3} \exp\left(\frac{-(q\sigma_{\rm surf})^2}{2}\right)\right]^2 + \frac{I(0)}{1+\xi^2 q^2},
\end{equation}
which depends on several adjustable fit parameters: the particle radius $R'$, the smearing parameter $\sigma_{\rm surf}$ which corresponds to about half the thickness of the fuzzy shell, the average correlation length of the PNIPAM network $\xi$ and the amplitude of the long wave-length contribution of the network fluctuations to the intensity $I(0)$\cite{fernandez2011microgel}.  
In simulations we directly calculate the form factors as 
\begin{equation}
P(q)=\frac{1}{N}\sum_{ij}\left\langle\exp(-\vec{q}\cdot\vec{r}_{ij})\right\rangle
\end{equation}
where the angular brackets indicate an average over different configurations. The results are reported in Fig.~\ref{fig:profiles_3.2}(b), showing a consistent decrease of the amplitude and number of oscillations as $Z$ increases, signalling that the internal structure becomes increasingly soft. Together with the numerical data, we also show the corresponding fits by Eq.~\ref{eq:fuzzy}, which provide clear evidence that the fuzzy sphere model fully captures the shape of the form factors for all values of the confining radius. 
We also report in the bottom-rightmost panel of Fig.~\ref{fig:profiles_3.2}(b) the form factor of the diamond-lattice microgel. We find that the latter well overlaps at low and high $q$ with the disordered microgel one generated for $Z=70\sigma$, but displays a different slope at intermediate $q$-values, being almost flat until $q\sigma \approx 0.8$, where a small peak appears. The latter comes from the crosslinkers, which are spatially correlated due to the underlying ordered structure of the network; indeed, the length scale associated to the position of the small peak roughly corresponds to the width of the oscillations observed in the density profile. The flat behaviour of $P(q)$ at slightly higher $q$-values is a consequence of such geometric correlations. Thus the diamond-lattice microgel is able to reproduce the fuzzy sphere shape for small wave vectors and to retain the self-similarity of the chains at higher $q$-values. However, it crucially fails to describe the network at intermediate scales due to its crystalline order.

According to the fuzzy sphere model, the normalized radial polymer density profile $\rho(r)/\rho_0$, where $\rho_0$ is the polymer density of the inner core, is\cite{stieger2004small}:
\begin{eqnarray}
\label{eq:profiles}
\frac{\rho(r)}{\rho_0}=
\begin{cases}
1 & {\rm if} \quad r < R_c \\[0.5em]
1 - \frac{(r - R' + 2\sigma_{\rm surf})^2}{(8\sigma_{\rm surf}^2)} & {\rm if} \quad R_c\leq r < R' \\[0.5em]
\frac{(R' - r + 2\sigma_{\rm surf})^2}{(8\sigma_{\rm surf}^2)} & {\rm if} \quad  R'\leq r< R'' \\[0.5em]
0  &{\rm if} \quad   r\geq R'',
\end{cases}
\end{eqnarray}
\noindent where $R_c=R'-2\sigma_{\rm surf}$ is the radius of the constant-density part of the particle, $R'$ is usually referred as the core radius and  $R''=R'+2\sigma_{\rm surf}$ is the total radius including the fuzzy shell. The latter is the estimate of particle radius that is commonly obtained from small-angle scattering experiments. 
From fitting the form factors, we extract the density profiles predicted by the model in Eq.~\ref{eq:profiles} and plot them as lines in Fig.~\ref{fig:profiles_3.2}(a) together with those calculated from simulations (symbols). Quite remarkably, we find an excellent agreement between the two. At very small distances, a small residual noise is still present in the numerical data and  averaging over different realizations is crucial to correct the $r\rightarrow 0$ behavior. We thus find confirmation that the fuzzy sphere model fully describes the microgel internal structure, providing an accurate estimate of the monomer density profiles.

\begin{figure}[h!]
\includegraphics[width=0.5\textwidth,clip]{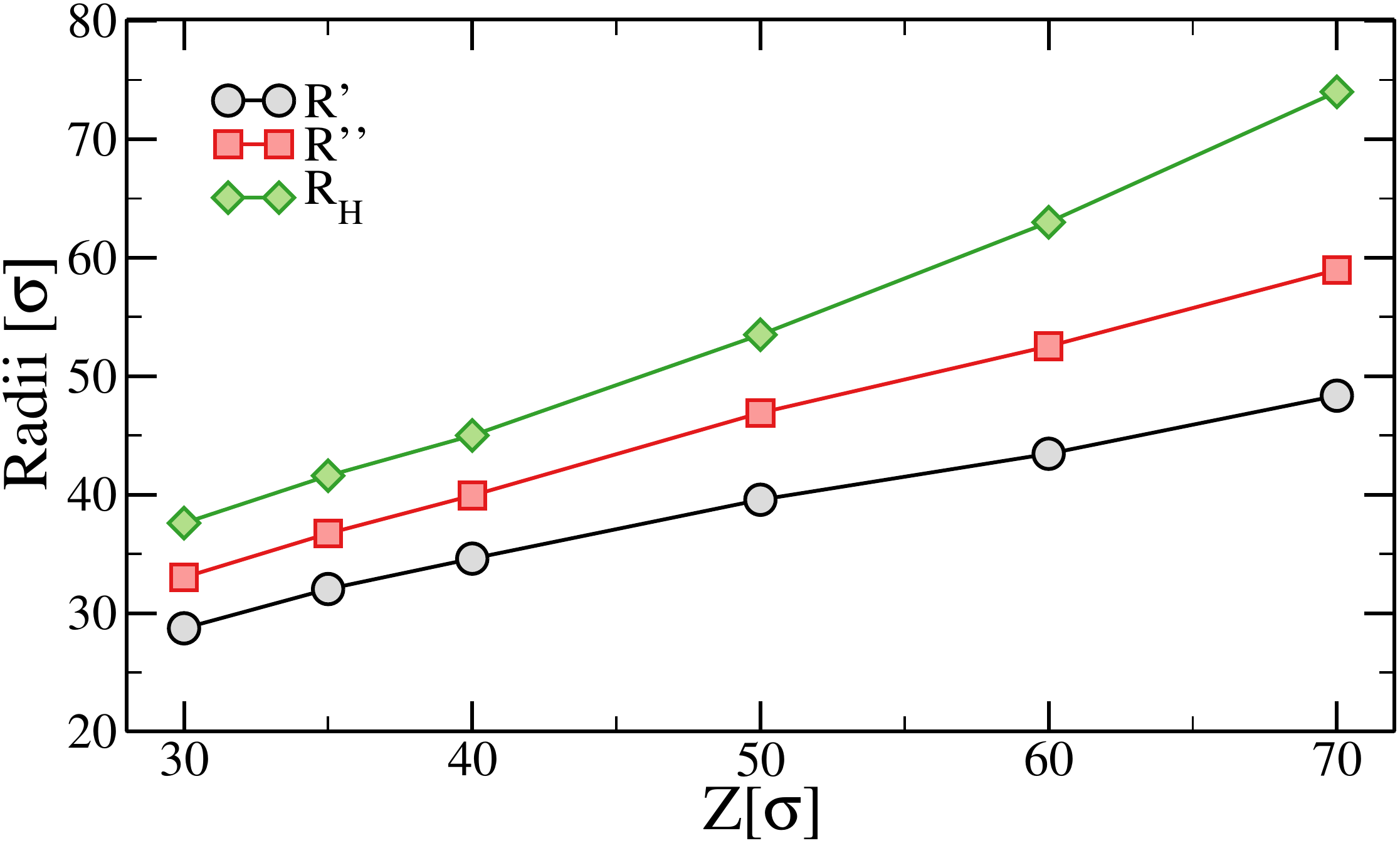}
\caption{\label{fig:fitparameters}Microgel radii obtained from fits to $P(q)$, $R'$ and $R''$, see Eq.~\ref{eq:profiles}, and directly extracted from simulations, $R_H$, for different values of the confinement $Z$.}
\end{figure}

In experiments the particle size obtained from the fits of the density profiles is commonly found to be systematically lower than the one estimated from Dynamic Light Scattering measurements, which is usually referred to as the hydrodynamic radius $R_H$. This systematic underestimation can be understood by looking at the density profiles close to the tail of the distributions. Indeed, as shown in the inset of Fig.~\ref{fig:profiles_3.2}(a), there is a residual tail, approximately exponential at large distances, which is not captured by the fuzzy sphere model. We adopt a qualitative threshold for determining the hydrodynamic radius: we define $R_H$ as the distance $r$ at which the normalized density profile becomes of the order of $10^{-3}$. The values for $R_H$ extracted in this way are plotted together with $R''$ and $R_c$ in Fig.~\ref{fig:fitparameters}. As expected, all radii increase monotonically with $Z$. 

\subsection{Temperature behaviour}
After having discussed the properties of the microgel in the swollen regime, we analyze the temperature behavior by setting $\alpha > 0$, thus in the presence of the solvophobic term $V_{\alpha}$.
After equilibration and for each value of the solvophobic parameter $\alpha$, we compute the swelling curves by calculating the ratio between the radius of gyration of the particles for a given $\alpha$, $R_g(\alpha)$, and its value at $\alpha=0$, that is, in the maximally swollen case, $R_g^{\rm MAX}$. The radius of gyration is directly calculated from simulations as $R_g=[1/N \sum_{i=1}^{N} ({\bf r}_i- {\bf r}_{\rm CM})^2]^{\frac{1}{2}}$ being ${\bf r}_{\rm CM}$ and ${\bf r}_i$ the position of the center of mass of the microgel and of the $i$-th bead, respectively. We have verified that $R_g$ and $R''$ are proportional to each other for all studied values of $Z$ and $\alpha$.
In order to compare experimental and numerical data, a relation between the parameter $\alpha$ and temperature must be established. To this aim, we scale the numerical data to match the volume phase transition temperature of experimental data, which, for PNIPAM microgels, is known to occur at $T \approx 307\,$K. Figure~\ref{fig:swelling} shows the resulting normalized radius of gyration $R_g/R_g^{MAX}$ as a function of temperature $T$, together with experimental results. 
{%\bf qua ovviamente andrebbe fatto con RH non Rg.... a sto punto...}
The comparison demonstrates that the solvophobic potential correctly captures the thermoresponsive behavior of microgels, allowing us to establish, for the present case, the linear relationship $T = 280.23\,$K $+ 38.33\,$K$\alpha$.
Fig.~\ref{fig:swelling} reports several swelling curves at different confining radii $Z$, showing that the change in $R_g$ increases by more than a factor of two if the microgel is generated in a sphere with radius $Z=70\sigma$ compared to the $Z=30\sigma$ case. This is a consequence of the looser structure of the microgel generated in the larger confining sphere which, consequently, has a larger available free volume in the swollen state compared to the one obtained with a smaller $Z$, which is more compact (see Fig.~\ref{fig:profiles_3.2}). Interestingly, the swelling curves for all confinements can be well fitted by the Flory-Rehner theory\cite{flory1953principles,lopez2007macroscopically,sierra2011swelling} which provides the equation of state of the microgel particle when the net osmotic pressure of the microgel is zero:

\begin{equation}\label{eq:Flory-Rehner}
\ln(1-\phi)+\phi+\chi(T,\phi)\phi^2+\frac{\phi_0}{N_{\rm gel}}\left[\frac{R_0}{R_g}-\frac{1}{2}\left(\frac{R_0}{R_g}\right)^3\right]=0.
\end{equation}

\noindent
 Here $\phi_0$ and $R_0$ are, the particle volume fraction and the radius of gyration in the collapsed state respectively, while $N_{\rm gel}$ is the average chain size, defined as the average number of monomers in a chain connecting two crosslinkers.
 In eq. (\ref{eq:Flory-Rehner}) $\chi(T,\phi)$ is a series expansion in $\phi$ of the Flory parameter, which also depends on the volume phase transition temperature $T_{VPT}$\cite{fernandez2011microgel}.
 %, which represents the free energy change when replacing  two solvent molecules in contact with a solvent-polymer contact. 
Inverting eq. (\ref{eq:Flory-Rehner}) an expression for the radius of gyration $R_g$  as a function of $T$ is obtained, that can be used as fitting function \cite{erman1986critical}.
 
% which accounts  for higher order interactions among molecules  being $\chi=\chi_1+\chi_2\phi+\chi_3\phi^2$. 
%An important role plays the Flory parameter which represents the free energy change when replacing  two solvent molecules in contact with a solvent-polymer contact $\chi_1=(1/2)-A(1-T_{VPT}/T)$. In the expression of $\chi_1$, $A$ is related to entropy variations while $T_{VPT}$ depends both on enthalpic and entropic changes and corresponds to  the temperature at which the system goes from a good solvent condition to a poor solvent condition; thus $T_{VPT}$ is the volume phase transition temperatue. Note that 

%\begin{equation}
%T_{\phi=0}=\frac{-A\phi_0^2(\frac{d_0}{d})^6 T_{VPT}}{\ln[1-\phi_0(\frac{d_0}{d})^3]+\phi_0(\frac{d_0}{d})^3+\phi_0^2(\frac{d_0}{d})^6 (\frac{1}{2}-A)+\frac{\phi_0}{N_{\rm gel}}[\frac{d_0}{d}-\frac{1}{2}(\frac{d_0}{d})^3] }
%\end{equation}
 
% In general, the fitting function depends on a large number of free parameters. We fix three of them: not only the VPT temperature to $307\,$K but also the radius  of gyration of the microgel particle in the collapsed state and the average chain length $N_{\rm gel}$, defined as the average number of monomers in a chain connecting two crosslinkers, to their values calculated in simulations. It turns out that $N_{\rm gel}$ is roughly constant and independent of $Z$, \textit{i.e.} $N_{\rm gel}=14$ for $c=3.2\%$. With these choices, the resulting curves, shown as solid lines in Fig.~\ref{fig:swelling}, are able to fully describe the numerical data.

In general, the fitting function depends on all the free parameters listed above.  We fix $T_{VPT}$ to $307\,$K as well as $R_0\simeq16.8$ and $N_{\rm gel}=14$, using the  ($Z$-independent) values of these quantities calculated in simulations. % as shown also in  Fig.~\ref{fig:chains_csd}.  
With these choices, the resulting curves, represented as lines in Fig.~\ref{fig:swelling}, are able to fully describe the numerical data.

\begin{figure}[h!]
\centering \includegraphics[width=0.5\textwidth,clip]{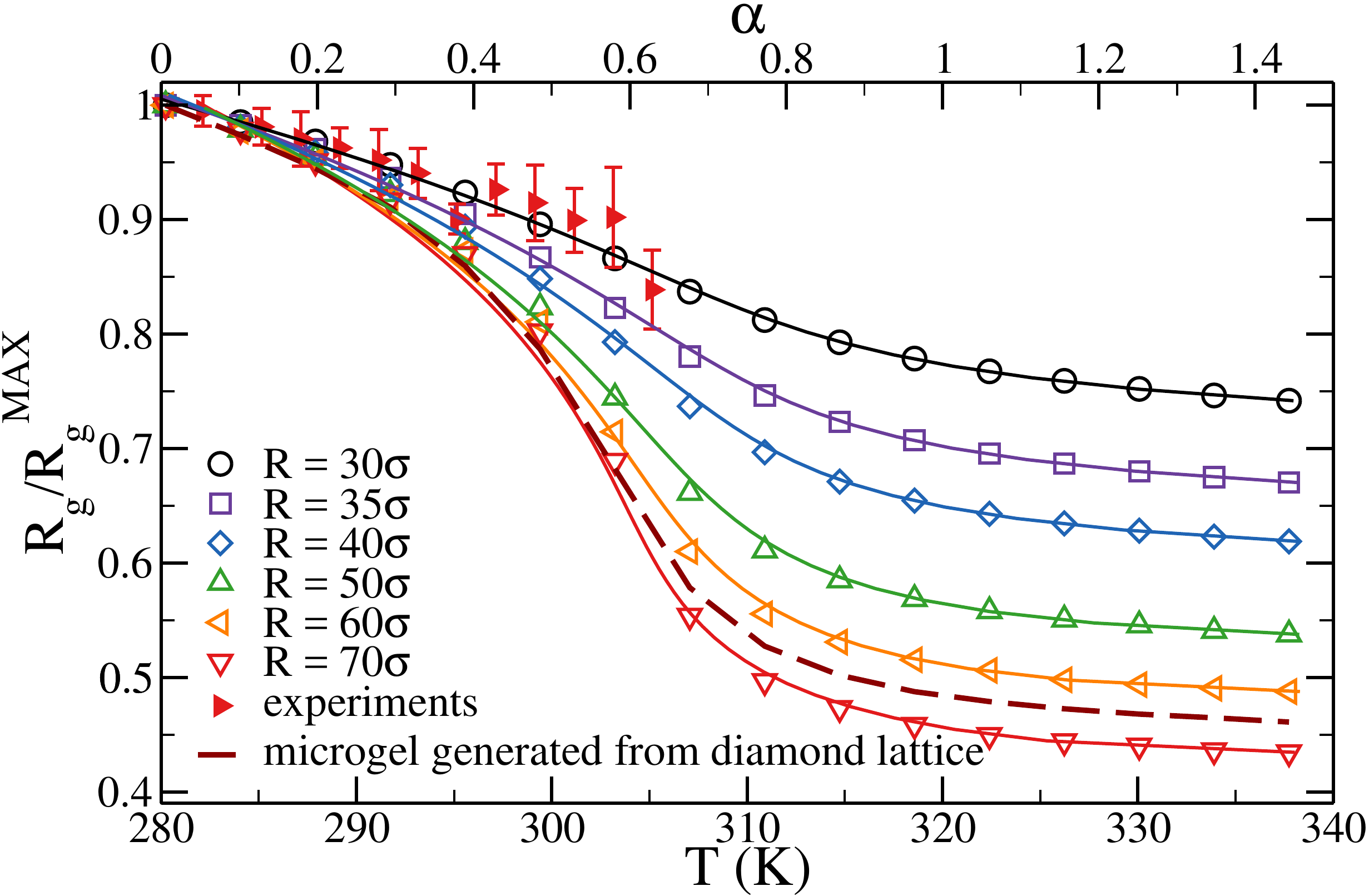}
\caption{Normalised radius of gyration $R_g/R_g^{\rm MAX}$ as a function of $\alpha$ (top axis) and temperature (bottom axis). Open symbols are numerical results obtained for microgels with $N\sim 41000$ monomer beads and a concentration of crosslinkers $=3.2\%$ for different values of the confining radius $Z$. Filled symbols are results from multi-angle DLS experiments; here the radius of gyration is normalized by its value at the lowest measured temperature, $T=282K$.  Solid lines are fits of the numerical data using the Flory-Rehner theory. The dashed line refers to the case of a microgel generated by a diamond lattice configuration.}
\label{fig:swelling}
\end{figure}

The microgels obtained for different confining radii display quite a distinct topology among themeselves. Even though $N_{\rm gel}$ is only weakly dependent on $Z$, a small confinement gives rise to a dense, possibly interwined, network, while large confinements allow particles to aggregate in a less intricate network due to the larger volume at disposal. We find that only the most strongly confined microgels ($Z=30\sigma$) obey the same small variation in size observed in the experimental data for small microgels. Thus $Z$ represents an useful control parameter that can be tuned to obtain microgels with different topologies and density profiles, while maintaining the same percentage of crosslinkers. This feature %{\color{red} of our synthesis protocol}
 is lost for instance for the case of a microgel with $N\simeq41000$ beads and $c=3.2\%$ generated from  a diamond lattice; the resulting curve is also reported in Fig.~\ref{fig:swelling} (dashed line). It is evident that these data, despite referring to a microgel having the same number of monomer beads and the same percentage of crosslinkers, cannot describe the experimental swelling curve. Indeed, the microgel obtained from the diamond lattice displays a swelling curve that is too steep compared to the small-microgel experimental data examined in this work. We further note that, since the topology of the microgel cannot be modified in the lattice synthesis approach, this is expected to also have important consequences on the mechanical properties at the single-particle level.
%By contrast, a quantitative agreement can be achieved with the disordered network microgel model that we have developed by appropriately tuning the confining radius. 

The successful comparison of the experimental swelling curve for small microgels with $c=3.2\%$ and that of our numerically generated ones for $Z=30\sigma$ provides us with an estimate of the bead size $\sigma$. Indeed, if we compare the experimental hydrodynamic radius at the lowest $T$, $R_H=24$~nm, with the numerical counterpart estimated for $\alpha=0$, i.e. $R_H\sim 37\sigma$, we obtain that our bead size is $\sigma\sim 0.65\,$nm. If $R''(\alpha=0)$ is used instead,  we would get a slightly larger value ($\sigma\approx 0.72\,$nm). Both values are consistent with the expected Kuhn length of PNIPAM polymers\cite{kutnyanszky2012there,magerl2015influence}. 
Thus, the choice to work with small microgels allowed us to avoid coarse-graining and to quantitatively match the number of crosslinkers in the simulations. However, we need to point out two drawbacks. The first one is that the experimental microgels are so small that their effective interaction, which can be approximated by a Hertzian model\cite{paloli2013fluid}, is very soft. Indeed, we estimated that at full overlap they do not exceed $\sim100 k_BT$\cite{Bergman17}. Hence, when brought above the VPT temperature, van der Waals attraction quickly takes over, leading to aggregation into large clusters. This is the reason for the lack of experimental data points for $T\gtrsim 305K$. Secondly, due to the small size of the microgels, SAXS measurements of the form factors have not been carried out, preventing a direct comparison with the numerical $P(q)$. Both these aspects will be tackled in the future by addressing the case of larger microgels with hydrodynamic radius of the order of a few hundreds nanometers. We note that in that case the monomer-resolved representation would require $\approx 10^7$ monomers. Thus, to undertake an investigation of such large particles, future approaches will need to be complemented by some coarse-graining procedure and by the analysis of system size dependence.

\begin{figure}[h!]
\hspace{-0.04cm}\includegraphics[width=0.97\textwidth,clip]{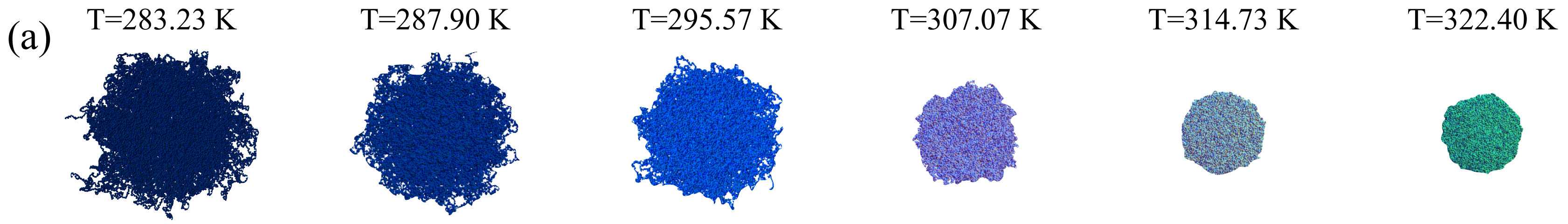}\\[0.2cm]
\includegraphics[width=0.48\textwidth,clip]{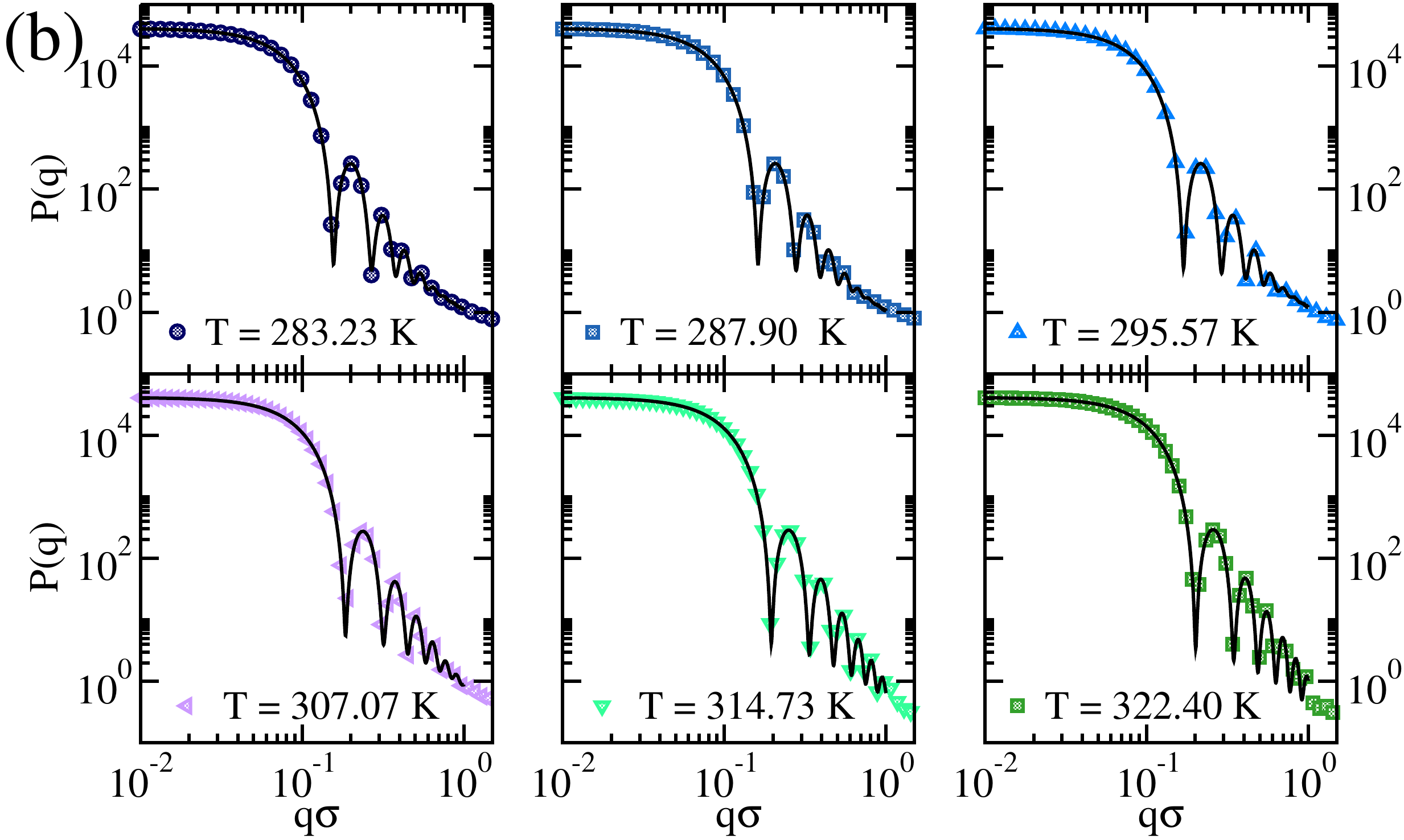}
\includegraphics[width=0.48\textwidth,clip]{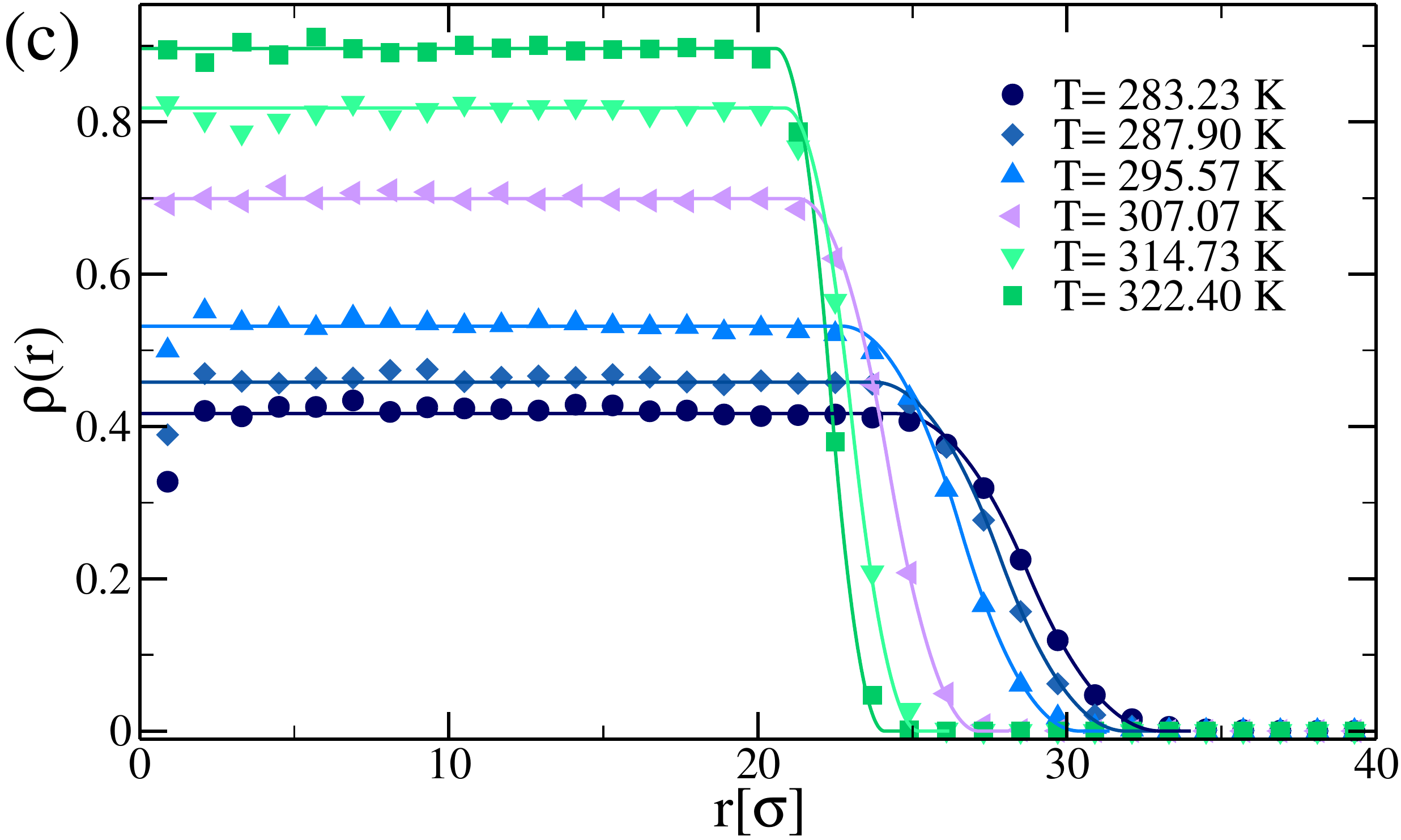}
\caption{(a) Snapshots of a microgel particle exhibiting the typical volume phase transition from swollen to compact. (b) Numerical form factors, averaged over four different realizations, for microgels with $N\sim 41000$ monomers and $c=3.2\%$  of crosslinkers generated in a sphere of radius $Z=30\sigma$ (symbols) at different values of the solvophobic parameter $\alpha$, corresponding to the reported temperatures as obtained from the swelling curve comparison to experiments in Fig.~\ref{fig:swelling}. Solid lines are fits of the curves using the fuzzy sphere model of Eq.~\eqref{eq:fuzzy}; (c) Averaged density profiles obtained from MD simulations of a microgel with $N\sim 41000$, $c = 3.2\%$ and $Z=30\sigma$ (symbols) and from the fit of the form factors using the fuzzy sphere model (solid lines).  }
\label{fig:snapshots_swelling}
\end{figure}

We have shown in Fig~\ref{fig:swelling} that the microgel particles generated in a confining sphere of radius $Z=30\sigma$ are the best candidates to reproduce the swelling properties of small microgels. We now focus on this value of $Z$ and investigate the behaviour of $P(q)$ as a function of $T$. The results are obtained after averaging over distinct microgel realizations. In Fig.~\ref{fig:snapshots_swelling}(b) we report $P(q)$  for different values of $T$. Similarly to what found in experimental form factors\cite{stieger2004thermoresponsive,richtering2011determination,virtanen2016persulfate,Paloliunpublished}, we find that, on increasing $T$, the peaks of $P(q)$ move to higher values of $q\sigma$, signalling a decrease of the size of the microgel. In addition, the number of peaks increases with $T$, a signature of the fact that the microgel becomes more and more compact. 
Snapshots of the corresponding microgels are shown in Fig.~\ref{fig:snapshots_swelling}(a), providing a visual confirmation of the occurrence of the VPT. The form factors are well described by the fuzzy sphere model of Eq.~\eqref{eq:fuzzy} at all $T$, mimicking the excellent agreement exhibited for the purely swollen case.
The density profiles extracted from the fuzzy sphere fits of $P(q)$ (Eq.~\eqref{eq:profiles}) are shown in Fig.~\ref{fig:snapshots_swelling}(c) and are compared with those calculated numerically. Once again, a very good agreement between the calculated and estimated density profiles is observed, confirming that the fuzzy sphere model is adequate to describe the microgel structure at all temperatures.

\section{Conclusions}
In this work we have introduced a novel approach to synthesize microgels {\it in silico}. We exploit the self-assembly properties of patchy particles to reproduce the polymerization of NIPAM monomers (two-patch particles) and BIS crosslinkers (four-patch particles). The two are fixed in concentrations by the molar ratio used in experiments. We specifically synthesized and measured the swelling curve of unusually small microgels, with hydrodynamic radius of the order of $24\,$nm in the swollen regime, in order to avoid complications due to coarse-graining, that will be tackled in the future.
By using an elegant method for achieving equilibration at very large attractive strengths~\cite{Sciortino2017}, we are able to build a fully-bonded network. While the kinetics of the network formation is clearly different from the experimental one, our aim is to obtain a particle which resembles the ones produced {\it in laboratory} as close as possible. We find that an important control parameter of our numerical synthesis is the volume in which particles initially form, and to this aim we confine the patchy particles in spheres of different radii. The confining radius $Z$ turns out to crucially control the network organization. Indeed, a strong confinement gives rise to a very intricate network with lots of entanglements, which is able to swell and deswell much less than a microgel generated in a larger volume. Once the network is formed, we preserve its topology by substituting the patchy model, useful to obtain the initial assembly, with a standard bead-spring polymer model. We further introduce the temperature dependence by adding a solvophobic term in the potential\cite{soddemann2001generic}.

Our computer-generated microgels are shown to share the main characteristics of their experimental counterparts. Their internal structure is characterized by a homogeneous core followed by a soft corona. The extension of the corona can be controlled in the preparation of the initial network by changing the confining radius. The microgel form factors are also shown to closely follow the fuzzy sphere model, widely used to describe the experimentally measured form factors. In addition, we have performed a consistency check for the density profiles, comparing the results of the direct calculations with those resulting from the fuzzy sphere assumption, finding perfect agreement for all investigated $Z$.

We next addressed the thermoresponsive properties of the microgels by comparing the swelling curve to the experimentally measured one for our test-case small microgels. We found that the experimental data can be well described only by microgels generated under strong confinement. This is likely due to the fact that the size of the experimental microgels is close to the lower limit imposed by the standard synthetic protocols. Thus, we expect that future comparisons with larger microgels will require less extreme confinements. Indeed, while the interpretation of the confinement is not straightforward in terms of the actual chemical synthesis, it surely correlates with the size of the resulting microgel. 
For the investigated experimental microgel we are able to reproduce the full temperature behavior and the occurrence of the volume phase transition, whose location is used to establish a relationship between the model solvophobic parameter $\alpha$ and the temperature. However a deeper investigation will be required also to assess how the chosen parameters in the hydrophobic interaction affect the swelling behavior of the microgel particles.
 Finally, the internal structure of the microgels at different $T$ is well-reproduced by the fuzzy sphere model, once again confirming the potential of our model to faithfully describe experimental data. The numerical protocol we developed was also shown to perform better than the ordered lattice approach used so far, which is not able to quantitatively reproduce the experimental swelling curve.

Unfortunately a direct comparison with experimental form factors for the investigated small microgels was not possible, due to the difficulty of the measurements. We will address this key issue in the near future by devising a coarse-graining strategy to tackle the assembly of microgels one order of magnitude larger in size (and thus roughly three orders of magnitude bigger in the number of particles). A careful design of the initial patchy particles and the conditions under which they will self-assemble will be crucial to achieve this goal. The present work is thus the first necessary step towards the more challenging task of designing a fully-fledged realistic microgel in the computer.

Our promising new approach to microgel computer simulations will also be crucial to evaluate effective interactions and to go beyond the widely used, but over-simplified, Hertzian model\cite{mohanty2014effective}. It will also allow us to carefully evaluate the role of dangling chains and dangling ends in determining the single-particle elastic properties as well as interparticle interactions. In addition, our \textit{in silico} synthesis protocol will make it possible to elucidate the important role of chain entanglements  in  the swelling behaviour as well as in the mechanical properties of microgels. Indeed, we have shown that their abundance in our particles is likely  controlled by the radius of the confining sphere used during the generation of the network. This is a topic that has recently gained a lot of interest, and a variety of methods\cite{bacova2017, caraglio2017physical} are being developed to evaluate entanglement effects, for which our model will provide an interesting test case.
Finally, starting from the neutral (at least in the swollen case) PNIPAM microgels case, we will extend the model by adding charges and, in a more ambitious project, to study interpenetrated networks microgels\cite{mattsson2009soft,nigro2015dynamic,nigro2017dynamical}, which have been shown to exhibit an intriguing fragile-to-strong transition~\cite{mattsson2009soft}, a very uncommon feature in colloidal systems. 

\section{acknowledgements}
NG and LR contributed equally to this work.
We thank P. Schurtenberger for comments on the manuscript and valuable help in the experimental part.
We also thank E. Chiessi, J.J. Crassous, F. Sciortino and L. Tavagnacco for fruitful discussions.
LR, NG and EZ acknowledge support from the European  Research Council (ERC Consolidator Grant 681597, MIMIC). 
MB acknowledges support from the European Research Council (ERC-339678-COMPASS).

\bibliographystyle{unsrt}
\bibliography{library}

\begin{thebibliography}{10}

\bibitem{vlassopoulos2014tunable}
Dimitris Vlassopoulos and Michel Cloitre.
\newblock Tunable rheology of dense soft deformable colloids.
\newblock {\em Current Opinion in Colloid \& Interface Science},
  19(6):561--574, 2014.

\bibitem{hansen2013theory}
Jean-Pierre Hansen and Ian~Ranald McDonald.
\newblock {\em Theory of simple liquids: with applications to soft matter}.
\newblock Academic Press, 2013.

\bibitem{auer2001prediction}
Stefan Auer and Daan Frenkel.
\newblock Prediction of absolute crystal-nucleation rate in hard-sphere
  colloids.
\newblock {\em Nature}, 409(6823):1020--1023, 2001.

\bibitem{pusey1986phase}
Peter~N Pusey and W~Van~Megen.
\newblock Phase behaviour of concentrated suspensions of nearly hard colloidal
  spheres.
\newblock {\em Nature}, 320(6060):340--342, 1986.

\bibitem{brambilla2009probing}
Giovanni Brambilla, Djamel El~Masri, Matteo Pierno, Ludovic Berthier, Luca
  Cipelletti, George Petekidis, and Andrew~B Schofield.
\newblock Probing the equilibrium dynamics of colloidal hard spheres above the
  mode-coupling glass transition.
\newblock {\em Physical review letters}, 102(8):085703, 2009.

\bibitem{zaccarelli2015polydispersity}
Emanuela Zaccarelli, Siobhan~M Liddle, and Wilson~CK Poon.
\newblock On polydispersity and the hard sphere glass transition.
\newblock {\em Soft Matter}, 11(2):324--330, 2015.

\bibitem{yunker2014physics}
Peter~J Yunker, Ke~Chen, Matthew~D Gratale, Matthew~A Lohr, Tim Still, and
  AG~Yodh.
\newblock Physics in ordered and disordered colloidal matter composed of poly
  (n-isopropylacrylamide) microgel particles.
\newblock {\em Reports on Progress in Physics}, 77(5):056601, 2014.

\bibitem{seth2006elastic}
Jyoti~R Seth, Michel Cloitre, and Roger~T Bonnecaze.
\newblock Elastic properties of soft particle pastes.
\newblock {\em Journal of Rheology}, 50(3):353--376, 2006.

\bibitem{mohanty2017interpenetration}
Priti~S Mohanty, Sofi N{\"o}jd, Kitty van Gruijthuijsen, J{\'e}r{\^o}me~J
  Crassous, Marc Obiols-Rabasa, Ralf Schweins, Anna Stradner, and Peter
  Schurtenberger.
\newblock Interpenetration of polymeric microgels at ultrahigh densities.
\newblock {\em Scientific Reports}, 7, 2017.

\bibitem{alsayed2005premelting}
Ahmed~M Alsayed, Mohammad~F Islam, Jian Zhang, Peter~J Collings, and Arjun~G
  Yodh.
\newblock Premelting at defects within bulk colloidal crystals.
\newblock {\em Science}, 309(5738):1207--1210, 2005.

\bibitem{han2008melting}
Yilong Han, NY~Ha, AM~Alsayed, and AG~Yodh.
\newblock Melting of two-dimensional tunable-diameter colloidal crystals.
\newblock {\em Physical Review E}, 77(4):041406, 2008.

\bibitem{iyer2009self}
Ashlee St~John Iyer and L~Andrew Lyon.
\newblock Self-healing colloidal crystals.
\newblock {\em Angewandte Chemie International Edition}, 48(25):4562--4566,
  2009.

\bibitem{debord2002color}
Justin~D Debord, Susan Eustis, S~Byul~Debord, Mark~T Lofye, and L~Andrew Lyon.
\newblock Color-tunable colloidal crystals from soft hydrogel nanoparticles.
\newblock {\em Advanced Materials}, 14(9):658--662, 2002.

\bibitem{pelton1986preparation}
RH~Pelton and P~Chibante.
\newblock Preparation of aqueous latices with n-isopropylacrylamide.
\newblock {\em Colloids and Surfaces}, 20(3):247--256, 1986.

\bibitem{saunders1999microgel}
Brian~R Saunders and Brian Vincent.
\newblock Microgel particles as model colloids: theory, properties and
  applications.
\newblock {\em Advances in colloid and interface science}, 80(1):1--25, 1999.

\bibitem{pelton2000temperature}
Robert Pelton.
\newblock Temperature-sensitive aqueous microgels.
\newblock {\em Advances in colloid and interface science}, 85(1):1--33, 2000.

\bibitem{oh2008development}
Jung~Kwon Oh, Ray Drumright, Daniel~J Siegwart, and Krzysztof Matyjaszewski.
\newblock The development of microgels/nanogels for drug delivery applications.
\newblock {\em Progress in Polymer Science}, 33(4):448--477, 2008.

\bibitem{fernandez2009gels}
Antonio Fern{\'a}ndez-Barbero, Iv{\'a}n~J Su{\'a}rez, B~Sierra-Mart{\'\i}n,
  A~Fern{\'a}ndez-Nieves, F~Javier de~las Nieves, Manuel Marquez,
  J~Rubio-Retama, and Enrique L{\'o}pez-Cabarcos.
\newblock Gels and microgels for nanotechnological applications.
\newblock {\em Advances in colloid and interface science}, 147:88--108, 2009.

\bibitem{lyon2012polymer}
L~Andrew Lyon and Alberto Fernandez-Nieves.
\newblock The polymer/colloid duality of microgel suspensions.
\newblock {\em Annual review of physical chemistry}, 63:25--43, 2012.

\bibitem{pich2010microgels}
Andrij Pich and Walter Richtering.
\newblock Microgels by precipitation polymerization: synthesis,
  characterization, and functionalization.
\newblock In {\em Chemical Design of Responsive Microgels}, pages 1--37.
  Springer, 2010.

\bibitem{fernandez2011microgel}
Alberto Fernandez-Nieves, Hans Wyss, Johan Mattsson, and David~A Weitz.
\newblock {\em Microgel suspensions: fundamentals and applications}.
\newblock John Wiley \& Sons, 2011.

\bibitem{sadowski2014intelligent}
Gabriele Sadowski and Walter Richtering.
\newblock {\em Intelligent hydrogels}, volume 140.
\newblock Springer, 2014.

\bibitem{likos2001effective}
Christos~N Likos.
\newblock Effective interactions in soft condensed matter physics.
\newblock {\em Physics Reports}, 348(4):267--439, 2001.

\bibitem{zhang2009thermal}
Zexin Zhang, Ning Xu, Daniel~TN Chen, Peter Yunker, Ahmed~M Alsayed, Kevin~B
  Aptowicz, Piotr Habdas, Andrea~J Liu, Sidney~R Nagel, and Arjun~G Yodh.
\newblock Thermal vestige of the zero-temperature jamming transition.
\newblock {\em Nature}, 459(7244):230, 2009.

\bibitem{pamies2009phase}
Josep~C P{\`a}mies, Angelo Cacciuto, and Daan Frenkel.
\newblock Phase diagram of hertzian spheres.
\newblock {\em The Journal of chemical physics}, 131(4):044514, 2009.

\bibitem{mohanty2014effective}
Priti~S Mohanty, Divya Paloli, J{\'e}r{\^o}me~J Crassous, Emanuela Zaccarelli,
  and Peter Schurtenberger.
\newblock Effective interactions between soft-repulsive colloids: Experiments,
  theory, and simulations.
\newblock {\em The Journal of chemical physics}, 140(9):094901, 2014.

\bibitem{denton2003counterion}
AR~Denton.
\newblock Counterion penetration and effective electrostatic interactions in
  solutions of polyelectrolyte stars and microgels.
\newblock {\em Physical Review E}, 67(1):011804, 2003.

\bibitem{likos2011structure}
Christos~N Likos.
\newblock Structure and thermodynamics of ionic microgels.
\newblock {\em Microgel Suspensions: Fundamentals and Applications}, pages
  163--193, 2011.

\bibitem{pelton2004unresolved}
Robert Pelton.
\newblock Unresolved issues in the preparation and characterization of
  thermoresponsive microgels.
\newblock In {\em Macromolecular Symposia}, volume 207, pages 57--66. Wiley
  Online Library, 2004.

\bibitem{rubinstein2010polymer}
Michael Rubinstein.
\newblock Polymer physicsthe ugly duckling story: Will polymer physics ever
  become a part of proper physics?
\newblock {\em Journal of Polymer Science Part B: Polymer Physics},
  48(24):2548--2551, 2010.

\bibitem{claudio2009comparison}
Gil~C Claudio, Kurt Kremer, and Christian Holm.
\newblock Comparison of a hydrogel model to the poisson--boltzmann cell model.
\newblock {\em The Journal of chemical physics}, 131(9):094903, 2009.

\bibitem{jha2011study}
Prateek~K Jha, Jos~W Zwanikken, Francois~A Detcheverry, Juan~J de~Pablo, and
  Monica~Olvera de~la Cruz.
\newblock Study of volume phase transitions in polymeric nanogels by
  theoretically informed coarse-grained simulations.
\newblock {\em Soft Matter}, 7(13):5965--5975, 2011.

\bibitem{kobayashi2014structure}
Hideki Kobayashi and Roland~G Winkler.
\newblock Structure of microgels with debye--h{\"u}ckel interactions.
\newblock {\em Polymers}, 6(5):1602--1617, 2014.

\bibitem{ghavami2016internal}
Ali Ghavami, Hideki Kobayashi, and Roland~G Winkler.
\newblock Internal dynamics of microgels: A mesoscale hydrodynamic simulation
  study.
\newblock {\em The Journal of chemical physics}, 145(24):244902, 2016.

\bibitem{Ahuali2017}
Silvia Ahualli, Alberto Mart{\'\i}n-Molina, Jos{\'e}~Alberto Maroto-Centeno,
  and Manuel Quesada-P{\'e}rez.
\newblock Interaction between ideal neutral nanogels: A monte carlo simulation
  study.
\newblock {\em Macromolecules}, (5)(50):2229--2238, 2017.

\bibitem{kobayashi2017polymer}
Hideki Kobayashi, Rene Halver, Godehard Sutmann, and Roland~G Winkler.
\newblock Polymer conformations in ionic microgels in the presence of salt:
  Theoretical and mesoscale simulation results.
\newblock {\em Polymers}, 9(1):15, 2017.

\bibitem{elvingson_jpcm}
Natasha Kamerlin and Christer Elvingson.
\newblock Tracer diffusion in a polymer gel: simulations of static and dynamic
  3d networks using spherical boundary conditions.
\newblock {\em Journal of Physics: Condensed Matter}, 28(47):475101, 2016.

\bibitem{elvingson_collapse}
Natasha Kamerlin and Christer Elvingson.
\newblock Collapse dynamics of core--shell nanogels.
\newblock {\em Macromolecules}, 49(15):5740--5749, 2016.

\bibitem{wca}
John~D. Weeks, David Chandler, and Hans~C. Andersen.
\newblock Role of repulsive forces in determining the equilibrium structure of
  simple liquids.
\newblock {\em The Journal of Chemical Physics}, 54(12):5237--5247, 1971.

\bibitem{Sciortino2017}
Francesco Sciortino.
\newblock Three-body potential for simulating bond swaps in molecular dynamics.
\newblock {\em The European Physical Journal E}, 40(1):3, 2017.

\bibitem{kutnyanszky2012there}
Edit Kutnyanszky, Anika Embrechts, Mark~A Hempenius, and G~Julius Vancso.
\newblock Is there a molecular signature of the lcst of single pnipam chains as
  measured by afm force spectroscopy?
\newblock {\em Chemical physics letters}, 535:126--130, 2012.

\bibitem{magerl2015influence}
David Magerl, Martine Philipp, Ezzeldin Metwalli, Philipp Gutfreund, Xing-Ping
  Qiu, Franc oise~M Winnik, and Peter Mu~ller Buschbaum.
\newblock Influence of confinement on the chain conformation of cyclic poly
  (n-isopropylacrylamide).
\newblock {\em ACS Macro Letters}, 4(12):1362--1365, 2015.

\bibitem{kremer1990dynamics}
Kurt Kremer and Gary~S Grest.
\newblock Dynamics of entangled linear polymer melts: A molecular-dynamics
  simulation.
\newblock {\em The Journal of Chemical Physics}, 92(8):5057--5086, 1990.

\bibitem{bernabei2011chain}
Marco Bernabei, Angel~J Moreno, Emanuela Zaccarelli, Francesco Sciortino, and
  Juan Colmenero.
\newblock Chain dynamics in nonentangled polymer melts: A first-principle
  approach for the role of intramolecular barriers.
\newblock {\em Soft Matter}, 7(4):1364--1368, 2011.

\bibitem{soddemann2001generic}
Th~Soddemann, Burkhard D{\"u}nweg, and Kurt Kremer.
\newblock A generic computer model for amphiphilic systems.
\newblock {\em The European Physical Journal E}, 6(1):409--419, 2001.

\bibitem{verso2015simulation}
Federica~Lo Verso, Jos{\'e}~A Pomposo, Juan Colmenero, and Angel~J Moreno.
\newblock Simulation guided design of globular single-chain nanoparticles by
  tuning the solvent quality.
\newblock {\em Soft Matter}, 11(7):1369--1375, 2015.

\bibitem{pelton2011microgels}
Robert Pelton and Todd Hoare.
\newblock Microgels and their synthesis: An introduction.
\newblock {\em Microgel Suspensions: Fundamentals and Applications}, 1:1--32,
  2011.

\bibitem{virtanen2016persulfate}
OLJ Virtanen, A~Mourran, PT~Pinard, and W~Richtering.
\newblock Persulfate initiated ultra-low cross-linked poly
  (n-isopropylacrylamide) microgels possess an unusual inverted cross-linking
  structure.
\newblock {\em Soft matter}, 12(17):3919--3928, 2016.

\bibitem{elaissari2011polymerization}
Abdelhamid Elaissari and Ali~Reza Mahdavian.
\newblock Polymerization kinetics of microgel particles.
\newblock {\em Microgel Suspensions: Fundamentals and Applications}, pages
  33--51, 2011.

\bibitem{hoare2006kinetic}
Todd Hoare and Daniel McLean.
\newblock Kinetic prediction of functional group distributions in
  thermosensitive microgels.
\newblock {\em The Journal of Physical Chemistry B}, 110(41):20327--20336,
  2006.

\bibitem{rey2016isostructural}
Marcel Rey, Miguel~{\'A}ngel Fern{\'a}ndez-Rodr{\'\i}guez, Mathias Steinacher,
  Laura Scheidegger, Karen Geisel, Walter Richtering, Todd~M Squires, and Lucio
  Isa.
\newblock Isostructural solid--solid phase transition in monolayers of soft
  core--shell particles at fluid interfaces: structure and mechanics.
\newblock {\em Soft Matter}, 12(15):3545--3557, 2016.

\bibitem{boon2017swelling}
Niels Boon and Peter Schurtenberger.
\newblock Swelling of micro-hydrogels with a crosslinker gradient.
\newblock {\em Physical Chemistry Chemical Physics}, 2017.

\bibitem{footnoteends}
We notice that our microgels are almost fully connected in the present protocol
  so that the resulting microgels have a negligible number of dangling {\it
  ends}. While we do not expect a significant difference between dangling ends
  and chains, we will address this issue in future work by adding one-patch
  particles to enhance the fraction of ends.

\bibitem{bianchi_FS}
Emanuela Bianchi, Piero Tartaglia, Emilia La~Nave, and Francesco Sciortino.
\newblock Fully solvable equilibrium self-assembly process:  fine-tuning the
  clusters size and the connectivity in patchy particle systems.
\newblock {\em J. Phys. Chem. B}, 111(40):11765--11769, 2007.
\newblock PMID: 17880197.

\bibitem{conley2016superresolution}
Gaurasundar~M Conley, Sofi N{\"o}jd, Marco Braibanti, Peter Schurtenberger, and
  Frank Scheffold.
\newblock Superresolution microscopy of the volume phase transition of pnipam
  microgels.
\newblock {\em Colloids and Surfaces A: Physicochemical and Engineering
  Aspects}, 499:18--23, 2016.

\bibitem{stieger2004small}
Markus Stieger, Walter Richtering, Jan~Skov Pedersen, and Peter Lindner.
\newblock Small-angle neutron scattering study of structural changes in
  temperature sensitive microgel colloids.
\newblock {\em The Journal of chemical physics}, 120(13):6197--6206, 2004.

\bibitem{flory1953principles}
Paul~J Flory.
\newblock {\em Principles of polymer chemistry}.
\newblock Cornell University Press, 1953.

\bibitem{lopez2007macroscopically}
T~Lopez-Leon and A~Fernandez-Nieves.
\newblock Macroscopically probing the entropic influence of ions: Deswelling
  neutral microgels with salt.
\newblock {\em Physical Review E}, 75(1):011801, 2007.

\bibitem{sierra2011swelling}
Benjamin Sierra-Martin, Juan~Jose Lietor-Santos, Antonio Fernandez-Barbero,
  Toan~T Nguyen, and Alberto Fernandez-Nieves.
\newblock Swelling thermodynamics of microgel particles.
\newblock {\em Microgel Suspensions: Fundamentals and Applications}, pages
  71--116, 2011.

\bibitem{erman1986critical}
B~Erman and PJ~Flory.
\newblock Critical phenomena and transitions in swollen polymer networks and in
  linear macromolecules.
\newblock {\em Macromolecules}, 19(9):2342--2353, 1986.

\bibitem{paloli2013fluid}
Divya Paloli, Priti~S Mohanty, J{\'e}r{\^o}me~J Crassous, Emanuela Zaccarelli,
  and Peter Schurtenberger.
\newblock Fluid--solid transitions in soft-repulsive colloids.
\newblock {\em Soft Matter}, 9(11):3000--3004, 2013.

\bibitem{Bergman17}
Maxime Bergman, Nicoletta Gnan, Marc Obiols-Rabasa, Janne-Mieke Meijer,
  Emanuela Zaccarelli, and Peter Schurtenberger.
\newblock Effective interaction potentials in binary mixtures of
  thermoresponsive microgels.
\newblock {\em to be submitted}, 2017.

\bibitem{stieger2004thermoresponsive}
Markus Stieger, Jan~Skov Pedersen, Peter Lindner, and Walter Richtering.
\newblock Are thermoresponsive microgels model systems for concentrated
  colloidal suspensions? a rheology and small-angle neutron scattering study.
\newblock {\em Langmuir}, 20(17):7283--7292, 2004.

\bibitem{richtering2011determination}
Walter Richtering, Ingo Berndt, and Jan~Skov Pedersen.
\newblock Determination of microgel structure by small-angle neutron
  scattering.
\newblock {\em Microgel Suspensions: Fundamentals and Applications}, pages
  117--132, 2011.

\bibitem{Paloliunpublished}
Divya Paloli, Jerome~J. Crassous, Priti Mohanty, Emanuela Zaccarelli, and Peter
  Schurtenberger.
\newblock to be submitted, 2017.

\bibitem{bacova2017}
Petra Ba{\v c}ov{\'a}, Federica Lo~Verso, Arantxa Arbe, Juan Colmenero,
  Jos{\'e}~A Pomposo, and Angel~J Moreno.
\newblock The role of the topological constraints in the chain dynamics in
  all-polymer nanocomposites.
\newblock {\em Macromolecules}, 50(4):1719--1731, 2017.

\bibitem{caraglio2017physical}
Michele Caraglio, Cristian Micheletti, and Enzo Orlandini.
\newblock Physical links: defining and detecting inter-chain entanglement.
\newblock {\em Scientific Reports}, 7, 2017.

\bibitem{mattsson2009soft}
Johan Mattsson, Hans~M Wyss, Alberto Fernandez-Nieves, Kunimasa Miyazaki,
  Zhibing Hu, David~R Reichman, and David~A Weitz.
\newblock Soft colloids make strong glasses.
\newblock {\em Nature}, 462(7269):83--86, 2009.

\bibitem{nigro2015dynamic}
Valentina Nigro, Roberta Angelini, Monica Bertoldo, Valter Castelvetro,
  Giancarlo Ruocco, and Barbara Ruzicka.
\newblock Dynamic light scattering study of temperature and ph sensitive
  colloidal microgels.
\newblock {\em Journal of Non-Crystalline Solids}, 407:361--366, 2015.

\bibitem{nigro2017dynamical}
Valentina Nigro, Roberta Angelini, Monica Bertoldo, Fabio Bruni,
  Maria~Antonietta Ricci, and Barbara Ruzicka.
\newblock Dynamical behavior of microgels of interpenetrated polymer networks.
\newblock {\em Soft Matter}, 2017.

\end{thebibliography}

\end{document}